\documentclass{PoS}

\usepackage{bm}        
\usepackage{amssymb}   
 \usepackage{amsmath}
\usepackage[square,comma,sort&compress,numbers]{natbib}

\newcommand*{\no}{\noindent}
\newcommand*{\bea}{\begin{eqnarray}}
\newcommand*{\eea}{\end{eqnarray}}
\newcommand*{\be}{\begin{equation}}
\newcommand*{\ee}{\end{equation}}

\newcommand*{\pref}[1]{(\ref{#1})}

\newcommand*{\nn}{\nonumber}

\newcommand*{\la}{\left\langle}
\newcommand*{\ra}{\right\rangle}
\newcommand*{\bma}{\begin{matrix}}
\newcommand*{\ema}{\end{matrix}}

\newcommand*{\op}{{\cal O}}

\title{Implications of strict gauge invariance for particle spectra and precision observables}

\ShortTitle{Gauge-invariant perturbation theory}

\author{\speaker{Axel Maas}\thanks{We are grateful to Ren\'e Sondenheimer for helpful discussions, and comments on the manuscript.}\\
        E-mail: \email{axel.maas@uni-graz.at}}
        
\author{Larissa Egger\\
	E-mail: \email{larissa.egger@uni-graz.at}
	\vspace{0.2cm}\\
        Institute of Physics, NAWI Graz, University of Graz, Universit\"atsplatz 5, A-8010 Graz, Austria\\
        }

\abstract{The discovery of the Higgs together with the excellent performance of the LHC allow to make precision tests of Brout-Englert-Higgs Physics, and especially its underlying field-theory. In this field theory strict gauge-invariance requires observable states to have a more involved structure than assumed in standard perturbation theory. This can lead to, likely rather very small, deviations in precision tests of the standard model. Here, the mechanism behind these deviations will be elucidated, and, as an example, its possible implications for the $R$ ratio at future linear colliders will be estimated.}

\FullConference{The European Physical Society Conference on High Energy Physics\\
		5-12 July, 2017\\
		Venice}

\begin{document}

It appears to be an almost trivial statement that physical states need to be gauge-invariant. Otherwise, results would depend on the choice of a gauge, and thus on a human decision.

However, gauge-invariance is field-theoretically much more than just BRST invariance. In particular, elementary fields like the Higgs or the electron are not gauge-invariant, and can thus not be physical \cite{'tHooft:1979bj,Banks:1979fi,Frohlich:1980gj}. This seems to yield a contradiction, as the description of experiments using them as physical degrees of freedom is essentially perfect \cite{pdg}.

The resolution lies in an intricate mechanism found by Fr\"ohlich, Morchio, and Strocchi (FMS) \cite{Frohlich:1980gj}, and the fact that the standard model is a very special theory \cite{Torek:2016ede}. An introduction to this topic can be found in \cite{Torek:2016ede}. It requires to first formulate all observables manifestly gauge-invariant. In electroweak physics, this requires the use of composite, i.\ e.\ bound state, operators. Using a prescription dubbed gauge-invariant perturbation theory it is then possible to evaluate these operators.

The simplest example is the scalar particle discovered at CERN, usually called the Higgs. A suitable gauge-invariant composite operator is $\op(x)=\phi^\dagger(x)\phi(x)$, where $\phi$ is the complex Higgs doublet. The FMS prescription \cite{Frohlich:1980gj,Torek:2016ede} is to fix to a suitable gauge, e.\ g.\ a 't Hooft gauge, and rewrite the Higgs field as usual as $\phi=v+\eta$, where $v$ is the vacuum expectation value. Expanding the correlator in the fluctuation field $\eta$ yields symbolically (precise formulas can be found in \cite{Torek:2016ede,Egger:2017tkd})
\be
\la\op^\dagger\op\ra=v^4+v^2\la\eta\eta\ra+v\la\eta\eta\eta\ra+\la\eta\eta\eta\eta\ra\approx v^4+v^2\la\eta\eta\ra+\op(v,\mathrm{couplings}^0)\label{higgs}.
\ee
\no So far, the first step is exact, but the individual terms on the right-hand side are not separately gauge-invariant, and only their sum is. The second step is now the decisive in the FMS mechanism: Expand the correlators in both the vacuum expectation value (or equivalently the fluctuation field) and the couplings. Then, the only remaining term is the usual Higgs propagator. This implies that to this order the gauge-invariant scalar should have the same pole structure, and thus the same mass, as the Higgs. Thereby, the physical states can be approximated by the gauge-dependent ones. This can be done for the whole standard model \cite{Frohlich:1980gj} and has in the Higgs-$W$/$Z$ system been confirmed in lattice simulations \cite{Maas:2012tj,Maas:2013aia,Torek:2016ede}.

This explains why standard perturbation theory works so well. But the FMS mechanism yields also two possible deviations. First, in theories which are structurally different such a one-to-one mapping is not necessarily possible \cite{Maas:2017xzh}. This has far-reaching consequences for BSM model building. At least in a toy model this has been confirmed in lattice simulations \cite{Torek:2016ede,Maas:2016ngo}. Secondly, the sub-leading contributions in \pref{higgs} lead to deviations of the left-hand-side from perturbation theory. These should even in the standard model be present. The aim here is to estimate their size.

For this note that the same dressing is necessary for fermions, as they carry a weak gauge charge. Thus, even an electron needs to be described by a gauge-invariant operator like \cite{Frohlich:1980gj,Egger:2017tkd}
\be
\mathcal{O}^{g}_f(x) = h^{\dagger}(x) f^g(x), \quad \mathrm{with} \qquad h = \begin{pmatrix} \phi_2^* & \phi_1 \\ -\phi_1^* & \phi_2 \end{pmatrix},\label{fermion}
\ee 
\no where $f$ is the fermion doublet and $g$ labels the generation. Like in \pref{higgs}, the corresponding propagator expands to the elementary fermion in leading order, and thus the composite state has the same mass as the elementary fermion \cite{Frohlich:1980gj,Egger:2017tkd}. In this course, the weak gauge charge, identified usually as flavor, is exchanged for a custodial charge. One of the custodial states corresponds then to, e.\ g., the electron and the other one to the neutrino.

Consider now a process like fermion production at a lepton collider. Denote the gauge-invariant composite states \pref{fermion} by capital letters, $E$ and $F$. It is then necessary to evaluate matrix elements like \cite{Maas:2012tj,Egger:2017tkd} ${\cal M}=\la E^+E^-\bar{F}F\ra$, i.\ e.\ bound-state--bound-state scattering.

\begin{figure}[t]
\includegraphics[width=\linewidth]{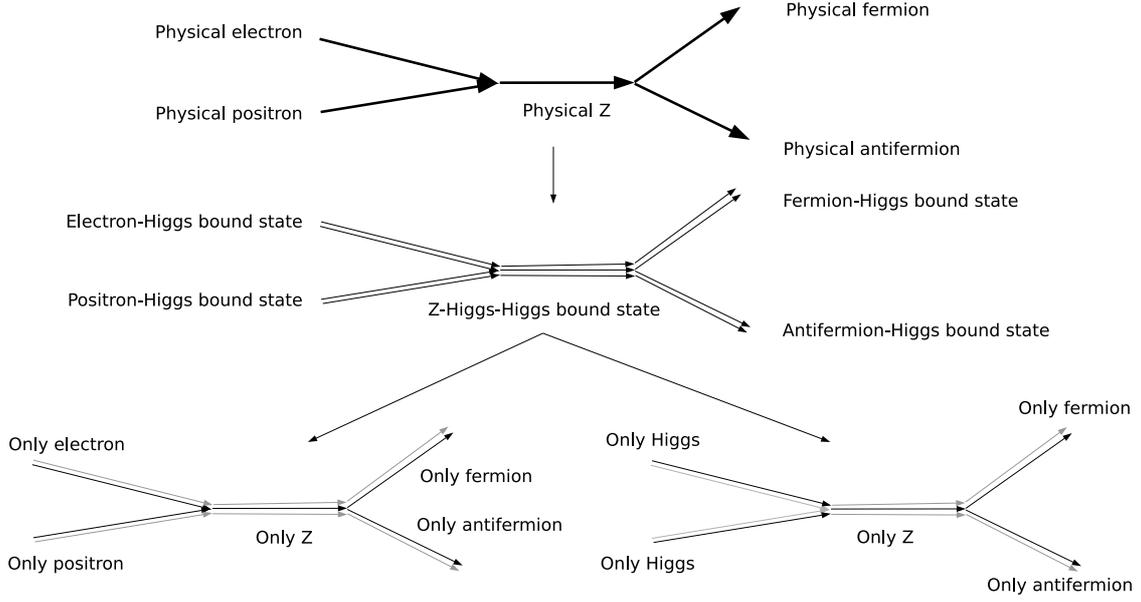}\\
\caption{\label{fig:scattering}Scattering process from the point of view of gauge invariance, for a sample intermediate state.}
\end{figure}

Applying the FMS prescription of \pref{higgs} and \pref{fermion} to the matrix element yields symbolically \cite{Egger:2017tkd}
\be
{\cal M} \approx v^4\la e^+e^-\bar{f}f\ra+v^2 \left( \la\eta^\dagger\eta\ra\la e^+e^-\bar{f}f\ra  +  \la e^+e^-\ra\la\eta^\dagger\eta\bar{f}f\ra  +  \la\bar{f}f\ra\la e^+e^-\eta^\dagger\eta\ra\right)\nn
+\mathrm{rest}
\ee
\no The leading term is the perturbative matrix element. The next-to-leading terms, suppressed by two powers of the vacuum expectation value, describe additional effects from the substructure. This is illustrated in figure \ref{fig:scattering}: Like in hadron-hadron scattering, the interaction is reduced to the interactions of the constituents. The other constituents play at this order the role of spectators. The first term in parentheses is a multiplicative modification of the perturbative matrix element. The second describes that the other constituent, the Higgs, interacts while the constituent electrons are spectators. The third describes the production of the additional constituents of the final states, and therefore corresponds to fragmentation. The rest are more complicated interactions at lower order in $v$. These subleading corrections will modify crosssections away from the perturbative, leading-order ones. But how much? A direct evaluation could miss effects due to the binding. 

\begin{figure}
\includegraphics[width=0.5\linewidth]{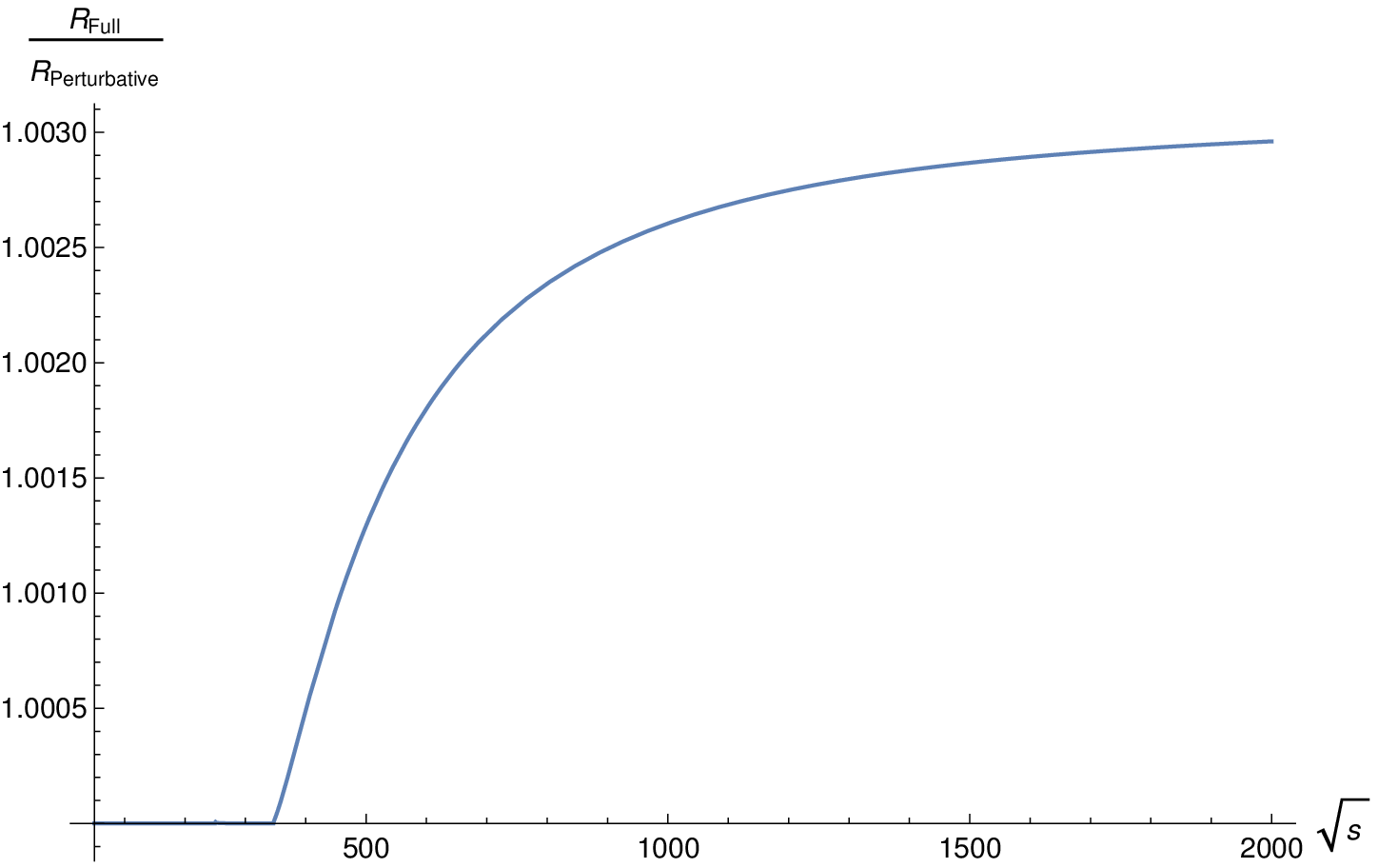}\includegraphics[width=0.5\linewidth]{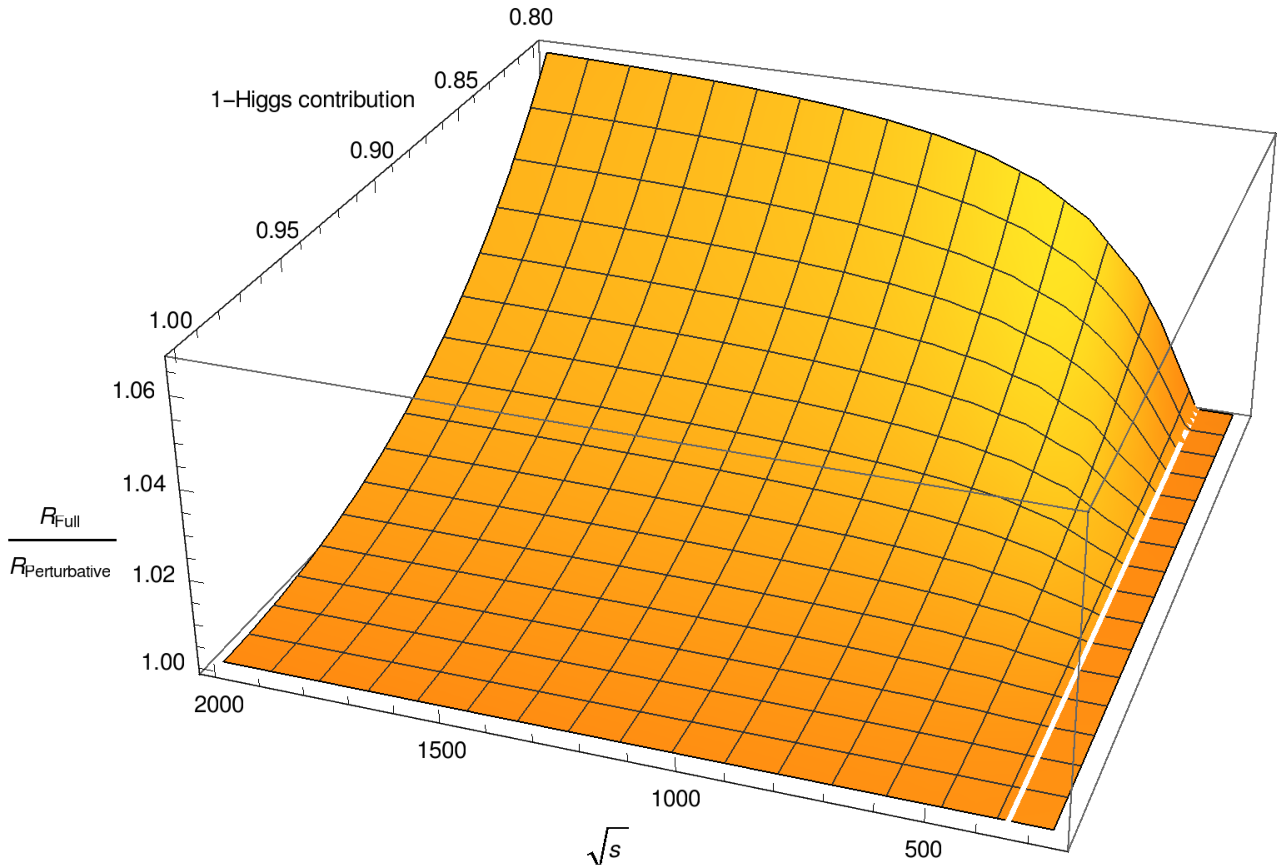}\\
\caption{\label{fig:rratio}The ratio of the full $R$ ratio to the perturbative one at rapidity zero as a function of $\sqrt{s}$ at a fixed Higgs contribution of $1-c=5$\% of the Higgs (left) and as a function of this contribution and $\sqrt{s}$ (right).}
\end{figure}

An alternative is to use a factorization ansatz, and introduce parton distribution functions and fragmentation functions for the constituents of the fermions in \pref{fermion}. This has been done in \cite{Egger:2017tkd}. There, it was found that the large mass of the Higgs makes effects below the Higgs threshold, and thus at LEP(2), unlikely. This explains why no effect has been seen so far. Unfortunately, this also makes the use of experimental data to constraint the PDFs impossible.

Making a model ansatz for the PDFs, in which the constituent electron carries most of the energy and the Higgs only a small fraction $1-c$ allows for an evaluation of the matrix element \cite{Egger:2017tkd}. For $\sigma_{e^+e^-\to\bar{f}{f}}$ this resulted essentially in a suppression of the crosssection by $c^2$ above the two-Higgs threshold, and thus a marked drop. However, this is not necessarily always so. Given this elementary cross-section, quantities like the $R$-ratio $\sigma_{e^+e^-\to\textrm{hadrons}}/\sigma_{e^+e^-\to\mu^+\mu^-}$ can also be calculated. This is shown in figure \ref{fig:rratio}. Here, the same contribution yields only an effect at the per mil level. Even for a sizable contribution of 20\%, the effect is not larger than 5\%. 

Concluding, field-theory predicts that due to gauge-invariance observable states should actually be composite. This implies that also in scattering processes sub-leading corrections should arise, which could be measurable, if the effect is large enough and a suitable observable is chosen. Understanding whether this prediction of field theory is correct is very important, as in BSM scenarios the same reasoning can lead to qualitative, rather than subtle, changes \cite{Torek:2016ede,Maas:2017xzh,Maas:2016ngo}.

\bibliographystyle{bibstyle}
\bibliography{bib}

\end{document}